# Stop Overthinking: Unlocking Efficient Listwise Reranking with Minimal Reasoning

Danyang Liu
*School of Computer Science and Technology*
*Beijing Institute of Technology*
Beijing, China
3120230887@bit.edu.cn

Kan Li
*School of Computer Science and Technology*
*Beijing Institute of Technology*
Beijing, China
likan@bit.edu.cn

***Abstract*—Listwise reranking utilizing Large Language Models (LLMs) has achieved state-of-the-art retrieval effectiveness. Recently, reasoning-enhanced models have further pushed these boundaries by employing Chain-of-Thought (CoT) to perform deep comparative analysis of candidate documents. However, this performance gain comes at a prohibitive computational cost, as models often generate thousands of reasoning tokens before producing a final ranking. In this work, we investigate the relationship between reasoning length and ranking quality, revealing an *overthinking* phenomenon where extended reasoning yields diminishing returns. To address this, we propose a *Length-Regularized Self-Distillation framework*. We synthesize a dataset by sampling diverse reasoning traces from a teacher model (Rank-K) and applying a Pareto-inspired filter to select traces that achieve high ranking performance with minimal token usage. By fine-tuning on these concise, high-quality rationales, the student model learns to internalize efficient reasoning patterns, effectively pruning redundant deliberation. Experiments on TREC Deep Learning and NeuCLIR benchmarks demonstrate that our method maintains the teacher's effectiveness while reducing inference token consumption by 34% – 37% across different retrieval settings, offering a practical solution for deploying reasoning-enhanced rerankers in latency-sensitive applications.***

## I. Introduction

Retrieval-augmented generation (RAG) has become a powerful paradigm in recent years, enabling large language models to leverage external knowledge retrieved from large corpora [1] In a typical RAG pipeline, the first stage retrieves a set of candidate documents (e.g., via BM25 [2] or dense bi-encoders [3]), which are then passed to a reranker to reorder them before the final generation step. Given the critical role of reranking in determining the quality of retrieved context, much attention has been devoted to improving this stage. While early works relied on simple pointwise scoring models, recent research has shifted toward more sophisticated strategies, including listwise and generative reranking [4], [5], [6], where the model considers the entire candidate set holistically and produces a ranked list directly.

A parallel trend in LLM research is the rise of reasoning-intensive models, such as OpenAI's o1 [7] and DeepSeek-R1 [8]. Inspired by this development, recent reranking models like Rank-K [9] incorporate explicit Chain-of-Thought (CoT) [10] reasoning during inference. Unlike standard rerankers that directly map inputs to a permutation, these models generate detailed intermediate reasoning — summarizing documents, comparing their relevance, and distinguishing subtle nuances among candidates—before producing the final ranking. This *inference-time reasoning* paradigm theoretically enhances the model's ability to handle complex or ambiguous queries by enabling deeper deliberation.

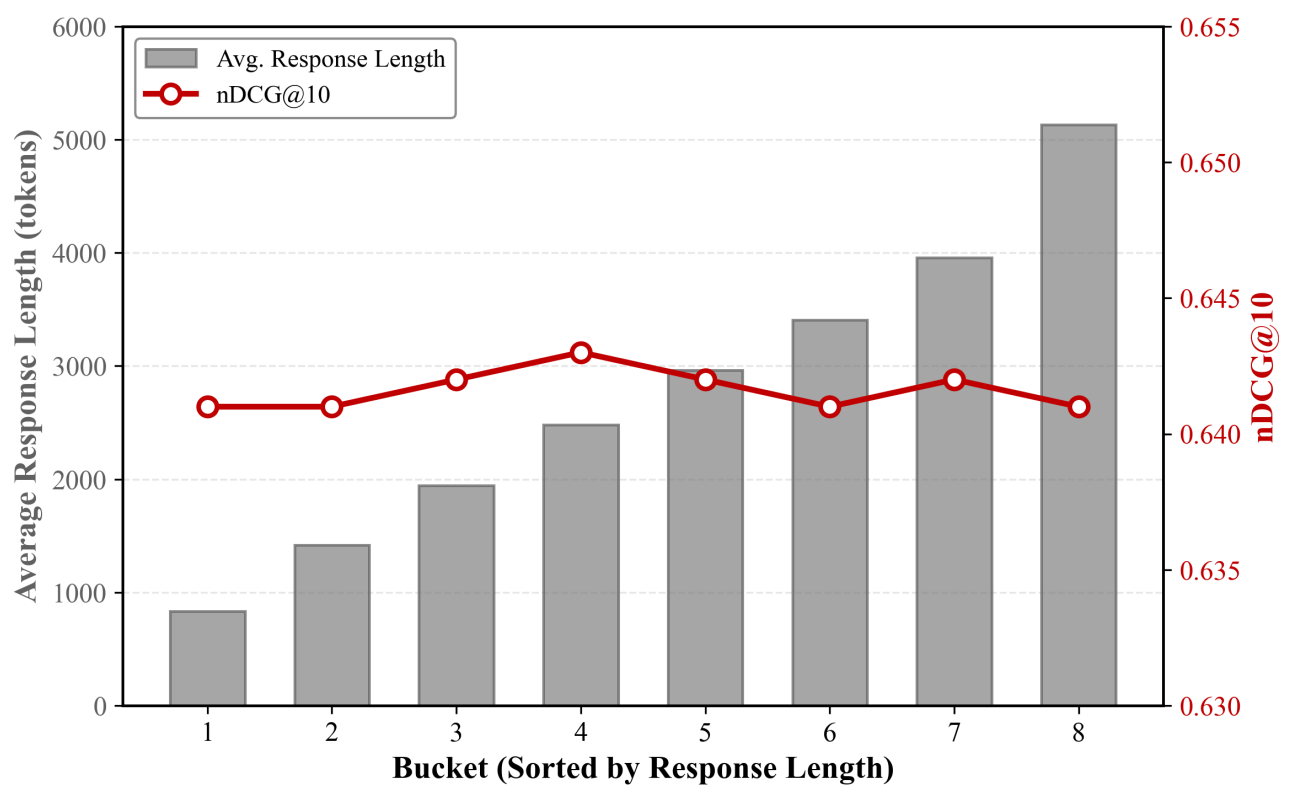

Fig. 1. Reasoning length vs. ranking performance on TREC DL20. Despite a six-fold increase in average response length (from 835 to 5,131 tokens), nDCG@10 remains stagnant around 0.642, revealing a clear overthinking phenomenon.

However, this reasoning capability introduces a critical bottleneck: *latency*. Reasoning-enhanced rerankers often generate thousands of tokens per query, rendering them impractical for real-time search applications where response time is paramount. This tension raises a fundamental research question: *Does extensive reasoning consistently translate to better ranking performance, or does the additional computation yield diminishing returns?* While scaling laws suggest benefits from extended computation [11], the specific trade-off between reasoning length and ranking quality remains underexplored.

In this work, we explore the trade-off between reasoning length and ranking effectiveness of generative rerankers. As shown in Fig. 1, we observe that **longer reasoning does not substantially improve ranking performance**. Based on this insight, we propose a self-distillation approach to derive efficient reasoning trajectories. We evaluate our method on TREC Deep Learning [12], [13] and NeuCLIR [14], [15], demonstrating that our approach reduces reasoning length by a substantial margin while maintaining competitive ranking accuracy. These findings offer a practical solution for deploying reasoning-enhanced rerankers in latency-sensitive applications.

To summarize, our main contributions are as follows:

- We identify and empirically demonstrate the *overthinking* phenomenon in reasoning-augmented rerankers, showing that extended reasoning yields diminishing returns for ranking quality.
- We propose a distillation-based approach that selectively learns from efficient reasoning trajectories, reducing reasoning overhead while preserving ranking performance.
- We conduct extensive experiments across multiple benchmarks and retrieval settings, demonstrating that our method generalizes well and offers a practical path toward efficient reasoning-enhanced reranking.

## II. Related Work

### A. Generative Listwise Reranking

Reranking serves as a critical refinement stage in information retrieval pipelines. While early neural approaches adopted pointwise architectures—scoring document-query pairs independently [16], [17], [18] —recent advancements have shifted towards listwise paradigms that model inter-document dependencies to capture relative relevance. The emergence of Large Language Models has accelerated this transition. Models such as RankGPT [4] leverage the instruction-following capabilities of LLMs to treat reranking as a generation task, directly outputting permuted document identifiers. Subsequent works, including RankZephyr [5] and RankVicuna [6] have further enhanced performance through specialized instruction tuning. Despite their effectiveness, standard generative rerankers typically map input contexts directly to rankings, which may limit their ability to handle complex queries that require multi-hop reasoning or fine-grained comparisons, thereby constraining their ultimate performance ceiling.

### B. Reasoning-Enhanced Retrieval

To address the limitations of direct ranking, recent research has integrated inference-time reasoning into IR, allowing models to generate explicit reasoning steps before predicting a rank. Rank-K [9] distills DeepSeek-R1 into a QwQ-32B based reranker that, given a query and a set of candidate passages, first reasons about each passage, groups and partially orders them, and then assembles a final ranking. This procedure is explicitly aligned with the Chain-of-Thought paradigm [10], and yields strong gains over prior listwise rerankers such as RankZephyr on TREC Deep Learning, NeuCLIR, and BRIGHT [19]. Similarly, Rank1 [20] explores distilling reasoning capabilities from proprietary reasoning models into smaller, open-source rerankers. While these methods achieve state-of-the-art performance, they introduce significant inference overhead. The generation of reasoning traces—often spanning thousands of tokens—transforms the latency bottleneck from input processing to output decoding. Prior studies have hinted at the inefficiency of these traces; for instance, Weller et al. [20] note that reasoning models can exhibit verbose or redundant patterns. Our observations align with this, suggesting that simply increasing the length of the reasoning trace does not guarantee performance improvements and may lead to diminishing returns in ranking metrics.

### C. Efficiency and Cost in Listwise Reranking

Efficiency remains a primary bottleneck for deploying listwise rerankers, primarily due to the quadratic or linear complexity relative to the list size and the limited context window of LLMs. To address this, prior works have proposed fixed-computation strategies. Sliding window approaches, adopted by RankGPT [4] and LRL [6], rerank overlapping subsets of candidates iteratively. Tournament-style methods, such as TourRank [21] and TDPart [22], partition candidates into groups for multi-round comparisons. More recently, AcuRank [23] introduced an uncertainty-aware adaptive computation framework, dynamically allocating reranker calls to ambiguous candidates rather than using fixed schedules.

However, these efficiency optimizations primarily focus on reducing the *number of inference calls* or the *input context length*. The emergence of reasoning rerankers introduces a new dimension of cost: the *decoding cost incurred by lengthy reasoning at inference time*. Since reasoning models like Rank-K generate long thought processes before outputting a ranking, the inference latency is dominated by the decoding time. Existing scheduling strategies do not address this redundancy within the generation process itself. Our work fills this gap by proposing a method to compress the reasoning trace, effectively reducing the *overthinking* phenomenon while maintaining the performance benefits of test-time reasoning.

## III. Methodology

We propose a ***Length-Regularized Self-Distillation*** framework that enables a reasoning-enhanced reranker to internalize concise yet effective reasoning patterns. The key idea is to leverage the model's own generation distribution: by sampling diverse reasoning trajectories and selectively fine-tuning on those that achieve strong rankings with minimal tokens, we encourage the model to learn efficient reasoning without sacrificing effectiveness.

### A. Problem Formulation

Let $q$ denote a user query and $\mathcal{D} = d_1, \dots, d_n$ be a set of candidate documents retrieved by a first-stage retriever, with associated relevance labels $\mathcal{R}_q$. A reasoning-enhanced listwise reranker with parameters $\theta$ generates a response $y$ that consists of a reasoning trace $r$ followed by a final ranking permutation $\pi$:

$$y = [r; \pi], \quad P_\theta(y \mid q, \mathcal{D}) = \prod_{t=1}^{|y|} P_\theta(y_t \mid y_{<t}, q, \mathcal{D}). \quad (1)$$

We measure ranking effectiveness using a metric $\mathcal{M}(\pi, \mathcal{R}_q)$ (e.g., nDCG@10) and denote the reasoning length by $|r|$. Ideally, we seek parameters that maximize ranking quality while minimizing reasoning cost:

$$\theta^\star = \underset{\theta}{\operatorname{argmax}}\, \mathbb{E}_{(q,\mathcal{D})}\big[\mathcal{M}(\pi, \mathcal{R}_q) - \lambda \cdot |r|\big], \quad (2)$$

where $\lambda > 0$ controls the trade-off between accuracy and efficiency. Direct optimization of (2) via reinforcement learning is unstable and computationally expensive. Instead, we approximate this objective through an efficiency-aware self-distillation procedure, where the model learns from its own high-quality, concise outputs.

## B. Efficiency-Aware Self-Distillation

Our approach is motivated by a key observation: for many inputs, the model's stochastic generation process already produces some trajectories that are both effective and concise. We refer to these as *efficient trajectories*. By identifying and distilling such trajectories back into the model, we can bias its generation distribution toward shorter reasoning without explicit length penalties. The procedure consists of three steps.

*1) Reasoning Trajectory Sampling:* We use Rank-K-32B [9] as the base model for both trajectory generation and subsequent fine-tuning. To avoid overlap with its original training data, we construct a seed query set $\mathcal{Q}_{\text{seed}}$ from MS MARCO v1, ensuring it is disjoint from the queries used in [9]. For each query $q \in \mathcal{Q}_{\text{seed}}$, we retrieve a candidate set $\mathcal{D}$ with $|\mathcal{D}| \in \{10,20\}$ using BM25.

To explore the model's reasoning space, we draw $K = 16$ stochastic samples per input:

$$\mathcal{S}_q = {y^{(k)}}_{k=1}^{K}, \quad y^{(k)} \sim P_\theta(\cdot \mid q, \mathcal{D}), \tag{3}$$

using nucleus sampling with temperature $\tau = 0.7, p = 0.95$, and maximum length $L_{\max} = 8192$ to encourage diversity in both content and length.

For each sample $y^{(k)}$, we parse the output to extract the predicted ranking $\pi^{(k)}$, compute its quality score $s^{(k)} = \text{nDCG@10}(\pi^{(k)}, \mathcal{R}_q)$, and record its token length $\ell^{(k)}$.

*2) Bicriteria Filtering:* Not all sampled trajectories are suitable for distillation. Some are short but inaccurate (under-thinking), while others are accurate but excessively verbose (over-thinking). We apply a bicriteria filter to identify trajectories that balance both objectives.

Let $\tilde{\mathcal{S}}_q \subseteq \mathcal{S}_q$ denote the subset of samples that (i) yield a valid, parseable ranking over $\mathcal{D}$ and (ii) achieve non-trivial performance ($s^{(k)} > 0$). We compute per-query statistics:

$$\mu_s = \frac{1}{|\tilde{\mathcal{S}}_q|} \sum_{y^{(k)} \in \tilde{\mathcal{S}}_q} s^{(k)}, \quad \mu_\ell = \frac{1}{|\tilde{\mathcal{S}}_q|} \sum_{y^{(k)} \in \tilde{\mathcal{S}}_q} \ell^{(k)}. \tag{4}$$

The *efficient candidate set* $\mathcal{E}_q$ contains samples that perform at or above the query-specific average while being shorter than average:

$$\mathcal{E}_q = \{ y^{(k)} \in \tilde{\mathcal{S}}_q \mid s^{(k)} \geq \mu_s \ \wedge \ \ell^{(k)} < \mu_\ell \}. \tag{5}$$

This simple rule excludes both under-thinking and over-thinking traces, biasing toward locally efficient trajectories. While (5) does not explicitly compute a Pareto frontier, it serves as a lightweight proxy that is effective in practice.

*3) Target Selection and Corpus Construction:* From each non-empty efficient set $\mathcal{E}_q$, we select the shortest trajectory as the distillation target:

$$y_q^\star = \underset{y^{(k)} \in \mathcal{E}_q}{\text{argmin}}\, \ell^{(k)}. \tag{6}$$

Queries for which $\mathcal{E}_q = \emptyset$ are discarded. This yields a distillation corpus:

$$\mathcal{T} = (q_i, \mathcal{D}_i, y_i^\star)_{i=1}^{N}, \tag{7}$$

containing concise, high-performing reasoning traces. In our experiments, we start with 50,000 seed queries; after filtering, approximately 22% are retained, yielding $N \approx 11{,}000$ training instances.

## C. Supervised Fine-Tuning

We fine-tune the same Rank-K-32B model on $\mathcal{T}$ using standard supervised learning. The training objective is the length-normalized negative log-likelihood:

$$\mathcal{L}(\theta) = -\frac{1}{|\mathcal{T}|} \sum_{(q,\mathcal{D},y^\star) \in \mathcal{T}} \frac{1}{|y^\star|} \sum_{t=1}^{|y^\star|} \log P_\theta(y_t^\star \mid y_{<t}^\star, q, \mathcal{D}). \tag{8}$$

Normalizing by $|y^\star|$ prevents the model from being biased toward shorter sequences purely due to fewer loss terms, ensuring stable optimization across varying output lengths.

By imitating these efficient trajectories, the model learns to preserve ranking effectiveness while implicitly pruning redundant reasoning steps, thereby mitigating the *overthinking* phenomenon.

# IV. Experimental Setup

## A. Datasets

We evaluate on two benchmark suites to assess both in-domain effectiveness and out-of-domain generalization. Table I summarizes the collection statistics.

TABLE I. Collection Statistics for TREC DL and NeuCLIR

| Dataset | | # Queries | # Docs | Avg. Len. |
|---|---|---|---|---|
| TREC DL | DL19 | 43 | 8,841,823 | 334.79 |
| | DL20 | 54 | | |
| NeuCLIR | Persian | 174 | 2,232,016 | 399.37 |
| | Russian | 171 | 4,627,541 | 307.91 |
| | Chinese | 167 | 3,179,206 | 372.52 |

*1) TREC Deep Learning*

We use the TREC Deep Learning 2019 and 2020 tracks [12], [13], the standard benchmarks for passage retrieval on the MS MARCO collection. These tracks provide dense, multi-graded relevance annotations suitable for fine-grained evaluation with nDCG@10.

*2) TREC NeuCLIR*

To assess generalization beyond MS MARCO, we employ the TREC NeuCLIR 2022–2024 tracks [14], [15]. Following standard practice for English-centric rerankers, we use the machine-translated English document collections (Farsi, Russian, Chinese) provided by the organizers. This setup tests whether our distilled model transfers to news articles with different stylistic and topical characteristics.

## B. Baselines

We compare our method (**Rank-K$_{\text{Distill}}$**) against the following systems:

*a) First-Stage Retrievers.* We use two retrievers to provide candidate lists of varying quality:

- **BM25** [2]: a standard lexical baseline.
- **SPLADE-v3** [24]: a state-of-the-art learned sparse retriever.

TABLE II. MAIN RESULTS ON TREC DL AND TREC NEUCLIR BENCHMARKS. WE REPORT NDCG@10 AND THE AVERAGE GENERATION LENGTH. THE BEST NDCG AND LOWEST LENGTH IN EACH BLOCK ARE BOLDED. '-' INDICATES NON-REASONING RERANKERS THAT OUTPUT ONLY A FIXED-LENGTH RANKING SEQUENCE.

| Retriever | Reranker | DL 2019 | | DL 2020 | | Farsi | | Russian | | Chinese | | Avg. | |
|---|---|---|---|---|---|---|---|---|---|---|---|---|---|
| | | nDCG | Len. | nDCG | Len. | nDCG | Len. | nDCG | Len. | nDCG | Len. | nDCG | Len. |
| BM25 | - | .499 | - | .485 | - | .386 | - | .337 | - | .375 | - | .416 | - |
| | RankZephyr | .645 | - | .633 | - | .291 | - | .279 | - | .274 | - | .424 | - |
| | Rank-K | .662 | 2284 | **.643** | 2606 | **.440** | 1872 | **.434** | 1873 | **.447** | 1863 | **.525** | 2100 |
| | Rank-$K_{Distill}$(Ours) | **.668** | **1379** | .641 | **1564** | .436 | **1123** | .429 | **1224** | .443 | **1318** | .523 | **1322**$_{\downarrow 37\%}$ |
| SPLADE-V3 | - | .722 | - | .751 | - | .507 | - | .449 | - | .428 | - | .571 | - |
| | RankZephyr | .775 | - | .804 | - | .398 | - | .406 | - | .314 | - | .539 | - |
| | Rank-K | .785 | 3184 | **.799** | 2835 | .581 | 2018 | **.528** | 2091 | **.518** | 2003 | **.642** | 2426 |
| | Rank-$K_{Distill}$(Ours) | **.795** | **2229** | .793 | **1616** | **.582** | **1227** | .516 | **1445** | .515 | **1520** | .640 | **1607**$_{\downarrow 34\%}$ |

*b) Reranking Baselines.*

- **RankZephyr** [5]: a 7B-parameter listwise reranker based on Mistral that directly outputs ranking permutations without intermediate reasoning.
- **Rank-K** [9]: a 32B-parameter reasoning-enhanced reranker that generates explicit Chain-of-Thought before ranking. This model serves as both our teacher and initialization, making the comparison critical for evaluating whether self-distillation preserves effectiveness while reducing reasoning overhead.

### C. Implementation Details

*a) Training Configuration.* We initialize from Rank-K-32B and perform full-parameter supervised fine-tuning on the distillation corpus $\mathcal{T}$. Training runs for 2 epochs with a global batch size of 64, using AdamW with learning rate $1 \times 10^{-5}$, $(\beta_1, \beta_2) = (0.9, 0.95)$, and weight decay 0.1.

*b) Inference Configuration.* At test time, we use nucleus sampling with temperature $\tau = 0.5$ and $p = 0.95$. We measure efficiency by the average number of tokens generated per query. All experiments are conducted on 8 NVIDIA H800 GPUs using the SWIFT framework.

## V. RESULTS AND ANALYSIS

### A. Main Results

Table II reports the performance of Rank-$K_{Distill}$ against baselines on TREC DL and NeuCLIR benchmarks. We analyze the results focusing on the trade-off between effectiveness and computational cost.

*1) Effectiveness-Efficiency Trade-off:* The primary finding is that **Rank-$K_{Distill}$ substantially reduces inference latency while preserving the ranking effectiveness of the teacher model.** Across all datasets, the student model achieves nDCG@10 scores comparable to, and occasionally surpassing, the original Rank-K model. For instance, on TREC DL 2019 (BM25), the distilled model outperforms the teacher (0.668 vs. 0.662). This suggests that our distillation process effectively filters out redundant or noisy reasoning paths that do not contribute to—or may even distract from—accurate relevance estimation.

Simultaneously, the efficiency gains are significant. Under the BM25 setting, Rank-$K_{Distill}$ reduces the average token generation by **37%**. Similarly, under the SPLADE setting, we observe a **34%** reduction. These reductions directly translate to lower latency and computational costs, addressing a critical bottleneck in deploying reasoning-enhanced rerankers.

*2) Robustness Across Domains:* Comparing against the non-reasoning baseline RankZephyr reveals the robustness of reasoning-based approaches. While RankZephyr performs well on TREC DL, it struggles significantly on the NeuCLIR benchmarks, often underperforming the first-stage retriever. In contrast, Rank-K $_{Distill}$ maintains strong performance across all domains, demonstrating that even with compressed reasoning, the model retains the ability to handle complex, diverse query distributions that challenge direct-mapping models.

### B. Analysis of Ranking Redundancy

To investigate the source of efficiency gains reported in Section 5, we analyze the generated reasoning traces both quantitatively and qualitatively. We posit that Rank-$K_{Distill}$ achieves efficiency not by simplifying the reasoning logic, but by pruning recursive self-correction loops and meta-cognitive overhead.

*1) Quantitative Metrics:* We introduce two metrics to quantify redundancy in the reasoning sequence $S = (R_1, \dots, R_T)$:

*a) Tail Repeat Ratio (TRR).* TRR measures the computational budget wasted after the model has effectively converged. Let $t^*$ be the index of the last unique ranking in $S$. TRR is defined as TRR $= (T - t^*)/T$. A high TRR indicates the model continues to generate tokens despite having already finalized its decision.

*b) Multi-Occurrence Ratio (MOR).* MOR quantifies the vacillation in decision-making. It is calculated as the ratio of unique rankings that appear more than once in $S$ to the total number of unique rankings. High MOR implies the model cyclically revisits previous states rather than progressively refining the ranking.

TABLE III. RANKING REDUNDANCY ANALYSIS ON TREC DL20.

| Model | Avg TRR | Avg MOR |
|---|---|---|
| Rank-K | 0.210 | 0.293 |
| **Rank-$K_{Distill}$ (Ours)** | **0.126** | **0.185** |

Table III confirms that the distilled model significantly reduces both metrics. This suggests that **Rank-K$_{\text{Distill}}$ learns to terminate generation promptly upon forming a stable ranking, minimizing unnecessary logical loops.**

*2) Qualitative Case Study*

To contextualize these metrics, Fig. 2 visualizes representative reasoning trajectories. The original Rank-K model exhibits a phenomenon we term *unproductive deliberation*: it constructs an initial ranking, subsequently expresses doubt about this decision (e.g., "*But need to check...*"), and after re-evaluation frequently returns to an essentially identical ordering.

Crucially, Rank-K$_{\text{Distill}}$ does not eliminate reflection entirely—doing so would risk overconfident errors. Instead, as shown in Fig. 2 (bottom), the distilled model proceeds with higher confidence once a stable ranking emerges, rather than repeatedly second-guessing settled decisions. This qualitative pattern supports our claim that self-distillation encourages the model to differentiate between *productive* reflection (which resolves genuine ambiguity) and *unproductive* deliberation (which merely restates already-finalized conclusions).

**Original Rank-K**

... *[Initial reasoning about passages]*
Overall, [13] directly states the filming location, [14] adds some contextual detail, and [19] mostly confirms the same fact.
**So the order would be:** `[13] > [14] > [19]...`
**But I should double-check** in case another passage is more specific...
*[Skims passages 3, 6, 7... but finds no new evidence]*
Maybe I undervalued [19]; it also confirms the same location.
**So maybe:** `[13] > [19] > [14]...`
**Wait, let me reconsider.** [14] mentions the specific studio lot, which is slightly more detailed than [19].
*[Re-reads 14 and 19, but refers to the same sentences as before]*
**So the final ranking would be:** `[13] > [14] > [19]...`
**But the user allows ties, so I could also group similar ones:** `[14] = [19]...`
**But the instructions prefer a strict ordering format, so I'll revert to:**
`[13] > [14] > [19] > [3] = [6] ...`

**Rank-K$_{\text{Distill}}$**

Passage [13] explicitly names the filming location, so it is the most relevant.

gives slightly more contextual detail about the same location, and [19] confirms the fact but adds no new information.
**Thus a natural ranking is:** `[13] > [14] > [19].`
**Wait, let me check.** none of the remaining passages provide a more precise location; they only repeat partial context.
Since ties are allowed, [14] and [19] could in principle be grouped, but [14] is marginally more informative.
**So the order above seems correct.**
`[13] > [14] > [19] > [3] = [6] = [7] ...`

Fig. 2. Comparison of reasoning traces. The **Original Rank-K** (top) repeatedly revisits settled decisions and slightly perturbs the ranking before reverting to its original choice, without introducing new evidence. Our **Rank-K$_{\text{Distill}}$** (bottom) performs a brief check but proceeds confidently once the ranking stabilizes, avoiding circular deliberation.

## VI. Conclusion

In this work, we addressed the latency bottleneck in reasoning-enhanced rerankers by identifying the phenomenon of *overthinking*, where excessive reasoning adds cost without improving ranking quality. To mitigate this, we proposed a *Length-Regularized Self-Distillation* framework. By distilling efficient trajectories from the model's own distribution, we enable it to internalize concise reasoning patterns. Experiments on TREC Deep Learning and NeuCLIR demonstrate that our approach reduces inference costs by over 30% while matching or exceeding the original model's performance.

## References

[1] S. Es, J. James, L. Espinosa Anke, and S. Schockaert, 'RAGAs: Automated Evaluation of Retrieval Augmented Generation', in Proceedings of the 18th Conference of the European Chapter of the Association for Computational Linguistics: System Demonstrations, N. Aletras and O. De Clercq, Eds., St. Julians, Malta: Association for Computational Linguistics, Mar. 2024, pp. 150–158. doi: 10.18653/v1/2024.eacl-demo.16.

[2] S. Robertson, H. Zaragoza, and others, 'The probabilistic relevance framework: BM25 and beyond', Found. Trends® Inf. Retr., vol. 3, no. 4, pp. 333–389, 2009.

[3] V. Karpukhin et al., 'Dense Passage Retrieval for Open-Domain Question Answering', in Proceedings of the 2020 Conference on Empirical Methods in Natural Language Processing (EMNLP), 2020, pp. 6769–6781.

[4] W. Sun et al., 'Is ChatGPT Good at Search? Investigating Large Language Models as Re-Ranking Agents', in Proceedings of the 2023 Conference on Empirical Methods in Natural Language Processing, 2023, pp. 14918–14937.

[5] R. Pradeep, S. Sharifymoghaddam, and J. Lin, 'RankZephyr: Effective and Robust Zero-Shot Listwise Reranking is a Breeze!', ArXiv Prepr. ArXiv231202724, 2023.

[6] R. Pradeep, S. Sharifymoghaddam, and J. Lin, 'RankVicuna: Zero-Shot Listwise Document Reranking with Open-Source Large Language Models', ArXiv Prepr. ArXiv230915088, 2023.

[7] A. Jaech et al., 'Openai o1 system card', ArXiv Prepr. ArXiv241216720, 2024.

[8] DeepSeek-AI et al., 'DeepSeek-R1: Incentivizing Reasoning Capability in LLMs via Reinforcement Learning'. 2025. [Online]. Available: https://arxiv.org/abs/2501.12948

[9] E. Yang et al., 'Rank-k: Test-time reasoning for listwise reranking', ArXiv Prepr. ArXiv250514432, 2025.

[10] J. Wei et al., 'Chain-of-thought prompting elicits reasoning in large language models', Adv. Neural Inf. Process. Syst., vol. 35, pp. 24824–24837, 2022.

[11] C. Snell, J. Lee, K. Xu, and A. Kumar, 'Scaling llm test-time compute optimally can be more effective than scaling model parameters', ArXiv Prepr. ArXiv240803314, 2024.

[12] N. Craswell, B. Mitra, E. Yilmaz, D. Campos, and E. M. Voorhees, 'Overview of the TREC 2019 deep learning track', ArXiv Prepr. ArXiv200307820, 2020.

[13] N. Craswell, B. Mitra, E. Yilmaz, and D. Campos, 'Overview of the TREC 2020 deep learning track. CoRR abs/2102.07662 (2021)', ArXiv Prepr. ArXiv210207662, 2021.

[14] D. Lawrie et al., 'Overview of the TREC 2022 NeuCLIR Track', in The Thirty-first Text REtrieval Conference (TREC 2022) Proceedings, 2023.

[15] D. Lawrie et al., 'Overview of the TREC 2023 NeuCLIR Track', in The Thirty-second Text REtrieval Conference (TREC 2023) Proceedings, 2024.

[16] R. Nogueira, W. Yang, K. Cho, and J. Lin, 'Multi-stage document ranking with BERT', ArXiv Prepr. ArXiv191014424, 2019.

[17] R. Nogueira, Z. Jiang, R. Pradeep, and J. Lin, 'Document Ranking with a Pretrained Sequence-to-Sequence Model', in Findings of the Association for Computational Linguistics: EMNLP 2020, T. Cohn, Y. He, and Y. Liu, Eds., Online: Association for Computational Linguistics, Nov. 2020, pp. 708–718. doi: 10.18653/v1/2020.findings-emnlp.63.

[18] R. Nogueira and K. Cho, 'Passage Re-ranking with BERT', ArXiv Prepr. ArXiv190104085, 2019.

[19] S. Hongjin et al., 'BRIGHT: A Realistic and Challenging Benchmark for Reasoning-Intensive Retrieval', in The Thirteenth International Conference on Learning Representations, 2024.

[20] O. Weller, K. Ricci, E. Yang, A. Yates, D. Lawrie, and B. Van Durme, 'Rank1: Test-time compute for reranking in information retrieval', ArXiv Prepr. ArXiv250218418, 2025.

[21] Y. Chen et al., 'TourRank: Utilizing Large Language Models for Documents Ranking with a Tournament-Inspired Strategy', ArXiv Prepr. ArXiv240611678, 2024.

[22] A. Parry, S. MacAvaney, and D. Ganguly, 'Top-Down Partitioning for Efficient List-Wise Ranking', ArXiv Prepr. ArXiv240514589, 2024.

[23] S. Yoon, G. Kim, G.-H. Cho, and S. Hwang, 'AcuRank: Uncertainty-Aware Adaptive Computation for Listwise Reranking', ArXiv Prepr. ArXiv250518512, 2025.

[24] C. Lassance, H. Déjean, T. Formal, and S. Clinchant, 'SPLADE-v3: New baselines for SPLADE', *ArXiv Prepr. ArXiv240306789*, 2024.